# On the "*Only Fields*" interpretation of Quantum Mechanics

Moses Fayngold
*Department of Physics, New Jersey Institute of Technology, Newark, NJ 07102*

The wave-particle duality, while being supported by overwhelming scientific evidence, is now claimed outdated by some physics professionals. They declare that the Quantum Field Theory allows only field aspect of reality. Presented below are the arguments showing that the Quantum Field Theory embraces the whole range of the Field-Particle Duality with both its extremes. We also discuss the important new experiments which reinforce the particle side of the story and thereby confirm the Field-Particle Duality.

## 1. Fields and particles

We revise the field-particle duality (FPD) (generalization of wave-particle duality), in the light of the quantum field theory (QFT). Such revision is relevant in view of the fashionable "*Only Fields*" (OF) interpretation of Quantum Mechanics (QM) shared by part of physics community [1-6]. The OF proclaims particles just the "excitations" of the respective fields. This caused counter-arguments [7-10] showing that QFT does not eliminate the concept of particles as sovereign entities and both – wave and particle "faces" – is a basic feature of a quantum object [11-14]. Presented below is, apart from analysis of FPD-QFT relationship, also a brief review of some new experiments demonstrating full validity of the FPD.

As a prelude, we start with definitions of particles and fields coming from Classical Physics (CPh) but remaining relevant today.

1) In CPh, a field is a spatial spread of a certain *directly measurable observable*, e.g. electric $\mathbf{E}(\mathbf{r},t)$ or magnetic $\mathbf{B}(\mathbf{r},t)$ field or their combinations including electromagnetic (EM) waves.

2) Particles are discrete and therefore countable; fields are continuous. A closely related property is that "*... particles are localized and waves are not*" [15]. In a position measurement, particle can be recorded as a single detector click or as a distinct spot on the observation screen.

3) A particle may generate one or more fields. This makes particle the primary object in CPh. Size and electric charge are the inner characteristics of a particle, not of a field.

4) A field generated by a particle is the tool for particle's interactions with the world.

The particle's *size* is not always uniquely defined. An electron's *effective* size $\simeq 2.4 \times 10^{-15} m$ is determined classically by its rest mass $9.109 \times 10^{-31} kg$ [11]. But in Particle Physics (PPh), the electron's *actual* size is considered zero since it has not been manifest in any known interactions. The PPh theorists call electron a point-particle. This implies that an electron charge acts on other charges but not on itself. The electron's rest mass still remains finite due to vacuum fluctuations producing a polarized cloud ("quantum foam") of virtual particles [16]. A nucleon size about $10^{-15}$ m is determined from the experiments, with about 5% range of discrepancy [17-19].

A field, on the other hand, cannot be ascribed a fixed size of its own. Already this alone disqualifies fields as the *primary* objects. Concept of field as an extended continuous entity does not seat well with fixed microscopic size of all its "*excitations*".

A photon has a special status in this argument. It is definitely not a point particle [20, 21]. On the other hand, it may be prepared with a sharply defined size, e.g. when released from a Fabry-Perot resonator (FPR). The common speed of all its spectral components preserves the state's form and size when propagating in a free space.

A photon released from FPR in two opposite directions is in a superposition of the two mutually receding forms; but each "half-self" retains its initial size. This size merely reflects the initial conditions (e.g., resonator's length), and in any case, a photon in its wholeness is recorded in one detector [22], which reveals its particle-like face, according to 2).

The field as defined in 3) can separate from the charge only partially. The *separated* part is never static and its origin can be traced back to its source. In classical limit, a change in the source's motion triggers field undulations eventually forming running waves *diverging* from the source. The energy-momentum of the "chipped-off" part of the field may come from some external agent accelerating the source. In all cases, the rest energy of the source is preserved. The released energy is eventually observed in the far-field (FF) domain as a superposition of diverging waves. The *basic* field of the source remains attached to it and can be observed in the near field (NF) domain [23]. The actual picture is more complicated and includes the reversed action on the source during the radiation process [23, 24].

The partial separation of EM field from its source promotes the classical field from an object's characteristic to the sovereign status. Then the emergence of QM in the early 20-th century unveiled the field-like properties of particles (superposition of local states) and particle-like properties of fields (energy quantization). This discovery of FPD has led to the concepts of *quantum particle* and *quantum field*.

With all outlined distinctions between particle and its field, there is no clear-cut borderline between them. The quantization of the field energy according to $\mathcal{E} = \hbar\omega$ has provoked some to consider any field as an ensemble of particles, and some others to consider all particles as localized quanta of the respective fields. As mentioned before, these two opposing views are just the extremes of general picture. Adhering solely to one of the extremes is an oversimplification throwing away the newly discovered richness of Reality [24, 25].

## 2. FPD from the standpoint of the QFT

Below the FPD is described from the viewpoint of QFT, and the analysis invalidates the OF.

The OF's proponents rightly replace the notion of classical field by *quantum field* $\Psi(\mathbf{r})$. But they fail to emphasize that $\Psi(\mathbf{r})$ is *not a directly measurable* observable. It can be measured only indirectly, mostly in a set of experiments on a pure ensemble of *particles*. Measurements reveal only probability density $\mathcal{P}(\mathbf{r}) = |\Psi(\mathbf{r})|^2$ but not the whole function $\Psi(\mathbf{r})$. Presenting $\Psi(\mathbf{r})$ as a superposition $\Psi(\mathbf{r}) = \sum c_j e^{i\alpha_j} \psi_j(\mathbf{r})$ of eigenstates $|\psi_j\rangle$ shows that full information about $\alpha_j$ necessary for determining the whole state is lost in $|\Psi(\mathbf{r})|^2$.

Einstein had rightfully dubbed $\Psi(\mathbf{r})$ a "Ghost Field". This term reflects a subtle nature of a quantum field as a superposition of virtual locations of a particle. Such superposition, forming a "probability cloud" in today's parlance, is the field-like face of a particle. This endorses the notion of *quantum field*. But at the same time, $\Psi(\mathbf{r})$ is the probability amplitude of finding *a particle* in vicinity of **r**. And we always find it in one piece at a *single* location. This shows the *particle* aspect as defined in 2), 3) of Sec.1. State $|\Psi\rangle$ can be written as $\Psi(\mathbf{r})$ only in the position basis $\Psi(\mathbf{r}) = \langle \mathbf{r} | \Psi \rangle$. Altogether, we have a *quantum particle* [12-14, 25].

Some OF-supporters argue that the process of "collapse" to a definite $\mathbf{r}'$ is a field property. But such an argument is based on a disastrously misleading interpretation of the term "collapse" as an instant convergence of some fluid-like substance to a small volume. Actual QM "collapse" has *nothing to do with it* and includes complementary "explosion" in the basis of an incompatible observable. This is a mathematical corollary of the QM indeterminacy. And both mutually opposing "jumps" can occur *instantly* only because $|\Psi\rangle$ is not directly measurable. The appropriate term for the whole phenomenon might be *Instant Reconfiguration* or *InstReguration* for brevity [25, 26]. (We cannot abbreviate it to IR which stands for Infra-Red). The *InstReguration* alone shows that a QM object is neither exactly field nor exactly particle but a much richer entity opening a far broader view of Nature.

The OF considers *position indeterminacy* $\Delta\mathbf{r}$ as invalidation of particles and ignores the object's *inner* size which shows its *particle-like* face. Another argument of the OF is: "*Field quanta have an all-or-nothing quality. Adding a monochromatic quantum to a field is nothing like adding a particle.*" [6].

This is like adding a *quantum particle*! An attempt to "chip off" a fraction of particle's energy at some location brings there the whole particle or nothing. A QM object propagates as a wave but is detected as a particle. Altogether, we have a *quantum field*! Both are in one!

Mathematically, the "Ghost Field" is a superposition of localized particle states $|\mathbf{r}\rangle$ according to

$$|\Psi\rangle = \int \chi(\mathbf{r})|\mathbf{r}\rangle d\mathbf{r} \qquad (1)$$

Here $|\mathbf{r}\rangle$ is an eigenstate of position operator $\hat{\mathbf{r}}$ [14, 24], and $\chi(\mathbf{r})$ is its local amplitude. Projecting (1) onto a state $|\mathbf{r}'\rangle$ gives position representation $\langle\mathbf{r}'|\Psi\rangle = \Psi(\mathbf{r}')$, $\langle\mathbf{r}'|\mathbf{r}\rangle = \delta(\mathbf{r}-\mathbf{r}')$ and $\chi(\mathbf{r}) = \Psi(\mathbf{r})$, so that (1) leads to familiar

$$\Psi(\mathbf{r}') = \langle\mathbf{r}'|\Psi\rangle = \int \Psi(\mathbf{r})\delta(\mathbf{r}-\mathbf{r}') d\mathbf{r} \qquad (2)$$

This is a textbook case of superposed *position eigenstates* in both – PPh and QFT. Relation (2) shows the intimate link between concepts of quantum particle and quantum field. And the InstReguration

$$\Psi(\mathbf{r}) \to \delta(\mathbf{r}-\mathbf{r}') \qquad (3)$$

just describes conversion from one "face" of an object to the other. Eq-s (1)-(3) form the basis of QFT, and neither of the opposing faces in them has the primary status.

The known arguments against legitimacy of operator $\hat{\mathbf{r}}$ and its eigenstates $|\mathbf{r}\rangle$ are that $\Delta\mathbf{r}\to 0$ is accompanied by unbounded increase of momentum and energy indeterminacy $\Delta\mathbf{p}$, $\Delta\mathcal{E}$. This leads to $\langle\mathcal{E}\rangle \to \infty$ with unrestricted pair production [27, 28].

Such arguments are unsubstantiated. The same logics could also "invalidate" operator $\hat{\mathbf{p}}$. Indeed, $\Delta\mathbf{p}\to 0$ leads to $\Delta\mathbf{r}\to\infty$, so the outcomes of practically all position measurements must lie beyond any finite domain of space. In addition, the states $|\mathbf{p}\rangle$ can reside only in pure vacuum without any boundaries and interactions. Such conditions are unrealistic, so states $|\mathbf{p}\rangle$ are not physically observable. And the same holds for energy eigenvalues $\mathcal{E}$ whose exact measurement requires infinite time and stationary conditions [23], which is also unrealistic. Thus, the eigenstates of continuous observables must, according to the OF, be all illegal.

But mathematical structure of QM and QFT embraces continuous variables by using the generalized normalization rule. Apart from $\langle\Psi_m|\Psi_n\rangle = \delta_{mn}$ for a discrete spectrum with integer $m$, $n$, we have

$$\langle\Psi_q|\Psi_{q'}\rangle = \delta(q-q') \qquad (4)$$

for continuous spectrum with an arbitrary observable $q$ [23 - 25]. Expressions (2), (3) are a special case of (4). Validity of eigenstates of observables with continuous spectrum is mathematically proved in [29].

Equal legitimacy of fields and particles can be expressed in the FPD language by naming a quantum object "*Wavicle*" as suggested by Eddington as far back as 1928 [30]; and now it can be updated to "*Fieldicle*". In the QFT, even stationary fields as defined in Sec.1 can be described in terms of *virtual particles*. The Coulomb field can be represented as longitudinally polarized *virtual photons* fleetingly emerging in vacuum fluctuations [3, 12-14].

The OF's basic claim that any particle is just an excitation of the underlying field is especially questionable for a stable particle. We do not see here any clear definition of the term "*Field excitation*". All known excitations are the Gamow states with finite lifetime. But an electron's lifetime is infinite, so it does not fall into this category. A proton lifetime exceeds $10^{34}$ years (about $10^{24}$ times longer than age of the universe) [31]. This is infinite for all practical purposes.

The same is true for a Hy atom in its ground state. Its $n$-th excited state has still finite radius $a_n = a_0 n^2$, so one can call it a concentrated state of electron field (field formulation) or a Gamow state of the electron particle (particle formulation). Altogether, we have a *Fieldicle*. That's what QFT is about.

The inconsistency of the *OF* formulation is seen in simple examples with photons. Historically photon was the first to be identified as a *particle* of the EM field due to its ability to excite an atom at a point **r**. But in contrast with an electron which preserves its identity within the atom, the absorbed photon just transmits its 4-momentum and spin to it and ceases to exist. In the QFT language, it InstRegurates to a vacuum state $|1\rangle \to |0\rangle$. But according to the OF, the resulting non-existing entity is just an excitation of the initial photon field! Conversely, the subsequent photon emission $|0\rangle \to |1\rangle$ is, in the OF language, an excitation of the previously non-existing photon field. Or consider the process

$$e^+ + e^- \to 2\gamma \tag{5}$$

Is the born photon pair the excitation of the pre-existing $(e^+, e^-)$ field or the excitation of the previously non-existing photon field? I have not found a persuasive answer to this question. Declaring statements without explanations does not look like a good description of reality.

Actually, the case (5) is an example of "*Transmutations between different kinds of particles*", which is an adequate formulation used in the PPh and QFT.

### 3. Some new experimental demonstrations of FPD

Of special interest may be a few recent experiments widely expanding the tested domains of FPD. One of them employs quantum entanglement [32]. The team used a pair of polarization entangled photons in a system that allows determining whereabouts of a tested photon within the field-particle spectrum. By manipulating one of the photons, the team managed to "*continuously morph, via entanglement, the test photon from wave to particle behavior even after it was detected. This result* illustrates the *inadequacy of a naive wave or particle description of light*."

Another experiment [33] shows how combined nanofabrication and nano-imaging methods allows recording the full build-up of quantum diffraction patterns in real-time for heavy molecules with a mass of 1298 amu. Wide-field fluorescence microscopy was used to detect the position of each molecule with an accuracy of 10 nm and to show the build-up of an interference pattern from stochastically arriving single molecules. Both – position detection of single molecules and incremental build-up of interference pattern by their stochastic arrivals are manifestations of their particle-like aspect.

One of the most recent works was performed with a linear array of the $^{87}$Rb atoms optically trapped in the nodes of EM standing wave across the optical axis of an FPR hosting two non-zero photon eigenstates [34]. The authors manipulated the atomic chain, shifting it along the resonator's axis. This allows to scan the optical field over a plane containing the FPR axis by detecting fluorescence from each atom and determining the accompanying Stark shift. According to the authors, "*A touchstone for super- resolution optical imaging ... is the precision with which they can resolve individual atoms, which are far smaller than the wavelength of the light used for imaging*.."

Quoting the review article [35]: "*Now, this problem ... has been turned on its head. Instead of using light to probe atoms, the atoms were used to probe an EM field.*" This has been done with resolution 2.5 times below wavelength.

In these experiments, "*...the atoms' tiny size*" (characteristic of a particle) is an additional factor showing full legitimacy of the particle "face" within the framework of QFT.

### 4. Conclusions

The FPD embraces both - *quantum fields* and *quantum particles* - as the opposite "faces" of quantum reality. The experiments with entangled photons and heavy molecules described in Sec.3 reinforce FPD and show the whole range of the field-particle spectrum. Generally, all available evidences up to now support FPD and show the new links and fascinating connections between field and particle aspects of reality.

Summarizing statement: a quantum object is neither exactly a particle nor exactly a field, but a far more complex and rich entity opening new dimensions beyond our classical intuition.

### Acknowledgements

I am deeply grateful to Art Hobson for his critical but constructive comments.I am deeply grateful to Art Hobson for his critical but constructive comments.